\documentclass[12pt]{article}
\usepackage[margin=1in]{geometry}  
\usepackage{chngcntr, mathtools, amsthm, amssymb, bbm, setspace, xcolor, arydshln, amsmath, amsfonts, array, comment, xcolor, hyperref}
\usepackage[utf8]{inputenc}
\usepackage[authoryear,round]{natbib}
\makeatletter
\renewcommand{\NAT@aysep}{}
\makeatother
\hypersetup{
    colorlinks,
    linkcolor={blue!50!black},
    citecolor={blue!50!black},
    urlcolor={blue!80!black}
}

\renewenvironment{thebibliography}[1]%
  {\begin{oldthebibliography}{#1}%
   \setlength{\itemsep}{0pt}%
   \setlength{\parsep}{0pt}%
   \setlength{\baselineskip}{0.9\baselineskip}} 
  {\end{oldthebibliography}}

\makeatletter
\g@addto@macro\normalsize{%
  \setlength\abovedisplayskip{5pt}   
  \setlength\belowdisplayskip{5pt}   
  \setlength\abovedisplayshortskip{0pt} 
  \setlength\belowdisplayshortskip{4pt}}
\makeatother

\usepackage{titlesec}  
\titleformat{\section}
  {\singlespacing\Large\bfseries}  
  {\thesection}{1em}{}

\makeatletter
\def\hline{%
	\noalign{\ifnum0=`}\fi\hrule \@height \arrayrulewidth \futurelet
	\@tempa\@xhline}
\makeatother

\title{\textbf{Gaussian Process Modeling with  \\ Genotype $\times$ Environment Kernels for Wheat Performance Prediction}}

\author{Lea Friedli\footnote{Engineering Risk Analysis Group, Technical University of Munich, Germany} , Tim Steinert\footnote{Institute of Mathematical Statistics and Actuarial Science, University of Bern, Switzerland}, Nathalie Wuyts\footnote{Agroscope, Plant-Production Systems, Switzerland}, Fabian Guignard\footnote{METAS, Federal Institute of Metrology, Switzerland},  \\ Lilia Levy H\"aner$^\ddagger$, Didier Pellet, Juan M. Herrera$^\ddagger$ and David Ginsbourger$^\dagger$}

\date{}

\begin{document}

\newcommand{\xg}{\boldsymbol{x}_G}
\newcommand{\xe}{\boldsymbol{x}_E}
\newcommand{\Xg}{\mathcal{X}_G}
\newcommand{\Xe}{\mathcal{X}_E}
\newcommand{\xge}{\boldsymbol{x}_{GE}}
\newcommand{\x}{\boldsymbol{x}}

\maketitle

\onehalfspacing
\begin{abstract} 
Optimizing wheat variety selection for high performance in different environmental conditions is critical for reliable food production and stable incomes for growers. We employ a statistical machine learning framework utilizing Gaussian Process (GP) models to capture the effects of genetic and environmental factors on wheat yield and protein content. In doing so, selecting suitable covariance kernels to account for the distinct characteristics of the information is essential. The GP approach is closely related to linear mixed-effect models for genotype $\times$ environment predictions, where random additive and interaction effects are modeled with covariance structures. However, while commonly used linear mixed effect models in plant breeding rely on Euclidean-based kernels, we also test kernels specifically designed for strings and time series. The resulting GP models are capable of competitively predicting outcomes for (1) new environmental conditions, and (2) new varieties, even in scenarios with little to no previous data for the new conditions or variety. While we focus on a wheat test case using a novel dataset collected in Switzerland, the GP approach presented here can be applied and extended to a wide range of agricultural applications and beyond, paving the way for improved decision-making and data acquisition strategies.
\end{abstract}


\section{Introduction}

Wheat is a staple crop essential for global food security, and its growth is influenced by an interplay of genetic, environmental, and management factors. Understanding how different wheat varieties respond to varying environmental conditions is crucial, especially in the face of climate change. When the goal is to identify genotypes optimal for specific environments, the patterns of genotype $\times$ environment \citep[G$\times$E;][]{kang1989genotype} interactions must be considered. Since not all combinations can be tested in the field, statistical and machine learning models serve as efficient surrogates \citep[e.g.,][]{fernandes2024using}.
In this paper, we consider Gaussian Process (GP) modeling for G$\times$E prediction, exploring various kernels defined on the G$\times$E product space. \\ 

We suppose that we observe some output values (potentially noisy) at certain input points, and our goal is to predict the output at new input points. In the context of G$\times$E predictions for plant selection, the output usually represents a performance trait, such as yield or grain protein content, and the inputs correspond to combinations of environments and varieties/genotypes (treated from here on as synonyms). We consider a setting in which we have additional information about the input combinations, as defined by the covariate tuple $\boldsymbol{x}$=($\xg$, $\xe$). For the environmental variables~$\xe \in \mathcal{X}_E$, we use meteorological variables that describe temperature, precipitation, and solar radiation, which are measured over six time periods within the wheat crop cycle \citep{costa2021nonlinear}. For the genetic information vectors~$\xg \in \mathcal{X}_G$, we consider single nucleotide polymorphisms \citep[SNPs; e.g.,][]{rafalski2002applications}. \\

One modeling approach which has been widely used in G$\times$E prediction is linear mixed-effect models 
\citep[LMMs; e.g.,][]{piepho1997analyzing, crossa2006modeling, herrera2018evaluation, buntaran2021projecting}.  
A baseline LMM for G$\times$E prediction models the response as the sum of an overall mean plus fixed and random effects due to the environment and/or the variety, plus an error term. Thereby, the random effects and errors are assumed to be Gaussian and independent, so no information is shared between different varieties or environments \citep{jarquin2014reaction}. However, a key advantage of a LMM is its ability to account for covariance structures, enabling strength to be borrowed across groups \citep{crossa2006modeling, burgueno2007modeling}. The numerous studies employing LMMs differ in their specification of fixed and random effects, as well as in how they handle interactions. Building on the overview provided by \citet{crossa2022genome}, we summarize the main modeling approaches used in the literature. \\

The basic single-environment genomic model analyzes genotypes within one environment, fitting a separate model for each. When genetic markers are available for the genotypes, the genetic effect is typically modeled using a parametric linear regression on the molecular markers. Assuming that the vector of marker effects follows an independent and identically distributed (i.i.d.) normal distribution leads to the ridge regression best linear unbiased prediction (rrBLUP) model. Alternatively, the genetic effect can be modeled using a centered multivariate Gaussian distribution, where the covariance matrix is modeled with a genomic relationship matrix derived from a kernel applied to the marker data. A specific choice of linear kernel yields the genomic best linear unbiased prediction (GBLUP) model \citep{vanraden2007genomic}.  
Other kernels, such as the Gaussian kernel, have been proposed to better capture non-linear marker effects and have shown improved performance in several studies \citep{cuevas2016genomic, cuevas2017bayesian, bandeira2017genomic}. More recently, kernels informed by deep learning techniques have also been explored \citep{cuevas2019deep, costa2021nonlinear}.\\

Multi-environment trials allow to borrow information across environments. One initial approach allows marker effects or genetic values to vary across environments \citep{schulz2013genomic, burgueno2012genomic}. However, such models typically enable prediction only within already observed environments. To enable prediction in new environments, environmental covariates can be incorporated to model dependencies among the random environmental effects, analogous to how genetic relationships are modeled using molecular markers. Models which specify genotype-specific variations due to key environmental factors were afterwards called reaction norm models \citep{crossa2022genome}. Capturing interactions between molecular markers and environmental covariates is challenging due to structural differences and potential high dimensionality. To address this, \citet{jarquin2014reaction} propose a modeling approach assuming that the interaction effect follows a normal distribution, with the covariance structure given by the element-wise (Hadamard) product of the two separate covariance matrices. 
\citet{cuevas2016genomic, cuevas2017bayesian} and \citet{bandeira2017genomic} demonstrated that incorporating a Gaussian kernel into multi-environment genomic models significantly enhances predictive performance compared to models employing a standard linear kernel. Building on this idea, \citet{costa2021nonlinear} extended the approach of \citet{jarquin2014reaction} by 
using both a Gaussian and a Deep kernel, within a Bayesian framework. \\

LMMs are typically fitted using (restricted) maximum likelihood and least square estimation. However, LMM methodology can also be reformulated in a Bayesian framework by treating all unknown quantities as random variables with assigned prior distributions \citep[e.g.,][]{meuwissen2001prediction}.
Bayesian inference then proceeds via the posterior distribution, allowing for joint estimation and full uncertainty quantification. Employing a full Bayesian framework, \citet{costa2021nonlinear} use a Gibbs sampler to generate the posterior distributions of the hyperparameters, effects and predictions. Recently, \citet{liu2025incorporating} employed a similar kernel-based model, where predictions are obtained through analytical formulas for Gaussian processes, following hyperparameter optimization based on Gibbs. In the work of both \citet{costa2021nonlinear} and \citet{liu2025incorporating}, SNP markers are numerically encoded, such that for both environmental and genetic information, Euclidean-based kernels are used. However, genomic data (e.g., SNPs) and environmental variables (e.g., time series from meteorological data) have distinct structures—strings and temporal signals, respectively. This motivates the use of more specialized kernels, here introduced within the framework of GP modeling. \\

\begin{figure}[p]
    \centering
    \includegraphics[width=\linewidth]{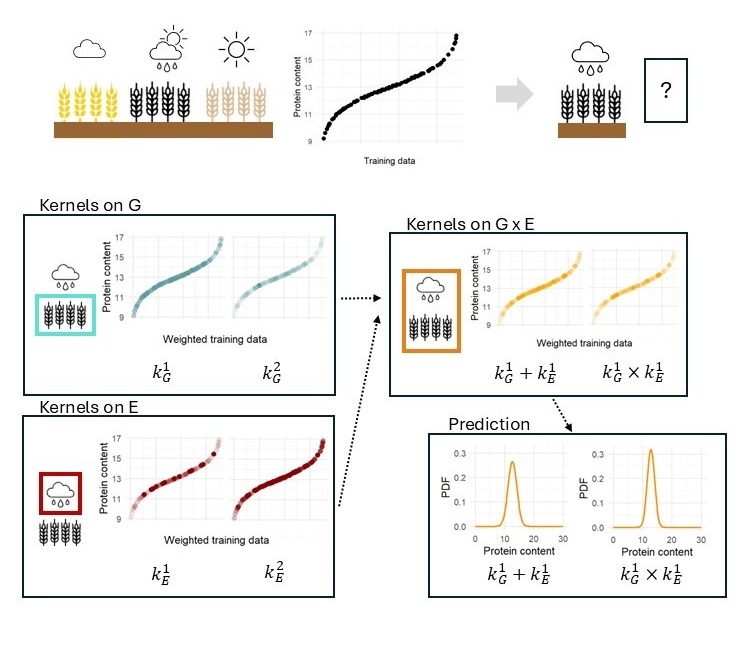}
    \caption{Illustration of GP modeling for G$\times$E prediction: Given training data from specific G$\times$E configurations, the goal is to predict outcomes (here the wheat grain protein content) for a new scenario. Kernels on $G$ and $E$ are used to quantify similarity between the training data and the target scenario, effectively weighting the training points (indicated by transparency of the colorful dots). The kernels are then combined to construct a joint kernel over the product space, and predictions are made based on the weighted training data according to this combined kernel. In GP modeling, the predictive distributions are Gaussian, as illustrated by the bell-shaped probability density functions depicted in prediction.  }
    \label{fig:illuGP}
\end{figure}

GP models originate from the geostatistical interpolation method known as kriging \citep{krige1951statistical}, but have broadened their application far beyond geostatistics \citep{rasmussen_williams}. GP modeling, aiming (in it its most common form) to learn a function \( f: \mathcal{X} \rightarrow \mathbb{R} \), can be applied to a wide range of input spaces $\mathcal{X}$, handling the diverse inputs through tailored covariance kernels. Apart from its versatility, GP modeling is popular for enabling uncertainty quantification, handling small training datasets, allowing the incorporation of prior knowledge through the covariance kernel, and providing interpretable results. In recent years, GPs have been increasingly applied across a variety of domains, including chemistry \citep[e.g.,][]{griffiths2023gauche}, engineering \citep[e.g.,][]{su2017gaussian} and environmental modeling \citep[e.g.,][]{ray2021bayesian}. 
In the considered GP approaches for G $\times$ E modeling, there are three major degrees of freedom illustrated in Figure~\ref{fig:illuGP}: the choice of the kernel for i) the genetic effects, ii) the environmental effects, and iii) the strategy used to construct a joint kernel over the product space of genotypes and environments.
Note: Here, “GP” refers to Gaussian process, not genomic-enabled prediction, which is also commonly abbreviated as “GP” in agricultural literature.\\

G $\times$ E kernels and associated GP models offer promising potential for decision-making and data acquisition strategies. Once a model is fitted, it provides probabilistic predictions at any candidate input, enabling the search for inputs that yield optimal responses.
As predictions are probabilistic, optimality would have to be defined in terms of the predictive distribution, be it in from a multi-objective perspective (e.g., in the spirit of Markowitz portfolio optimization) or via a single criterion such as a predictive quantile at a specified level \citep[e.g.,][]{picheny2013quantile}.
GP models have been used for global optimization for several decades and the resulting field now referred to as Bayesian Optimization (BO) has become very active from engineering to machine learning and beyond \citep{movckus1974bayesian, jones1998efficient, garnett2023bayesian}. 
GP-based modeling often separates controllable from non-controllable variables, aiming to optimize performance over the controllable ones under average or worst-case conditions \citep{williams2000sequential, janusevskis2013simultaneous, ginsbourger2014bayesian}. \\

In optimizing agronomic traits, one immediate application of GPs is using predictive distributions to compare and select varieties for specific environments.
Identifying the best genotype among a candidate population has also been addressed in the realm of LMMs in \citet{tanaka2018bayesian} and \citet{tsai2021bayesian}. However, environmental conditions have been ignored, and since the exact conditions during the target period may be unknown at the time of variety selection, GP methods are particularly promising for handling the associated uncertainties. 
Apart from genotype selection, GP models may be leveraged to design novel agronomical experiments. Experimental design in the G $\times$ E context has also been targeted using LMMs, for instance to optimize the allocation of trials to sub-regions in \citet{prus2024optimizing}.  
Here, the selection strategy aims to optimally predict the pairwise linear contrasts of the genotypes among some pre-defined sub-regions. In comparison, GPs provide a very flexible framework for any input variable space. Furthermore, GP models allow to work out acquisition functions dedicated to targeted experimental design, e.g. focusing on specific ranges of the response \citep{chevalier2014fast}.   \\

GP regression is closely related to LMM approaches; however, GPs typically rely on maximum likelihood estimation for hyperparameters and use analytical expressions for predictions, without requiring sampling-based inference (see the analytical Gaussian predictions in Fig. \ref{fig:illuGP}). Given the potential of GPs in the context of decision-making and data acquisition strategies, our aim is to bridge the gap between the methods and pave the way for promising future applications. Furthermore, while the state-of-the-art LMMs for G$\times$E prediction rely on Euclidean-based kernels for both environmental and genetic inputs, we also test kernels specifically designed for strings and time series. While \citet{costa2021nonlinear} investigate a test case involving maize hybrids, this manuscript analyzes a novel dataset targeting wheat yield and grain protein content, collected between 1990 and 2023 in multi-environment trials across the Swiss wheat production zone \citep{wheatdata}. 
The manuscript is organized as follows: Section~\ref{sec_gpmod} reviews GP modeling, G$\times$E kernels, and links to LMMs. Section~\ref{sec_dataimple} describes the dataset, implementation, metrics, and baselines. Finally, Section~\ref{sec_results} presents the wheat test case, and Section~\ref{sec_discu} concludes with discussion and summary.

\section{Gaussian Process Regression with combinations of Genotype and Environment Kernels}
\label{sec_gpmod}

\subsection{Gaussian Process basics}
\label{GPmodeling}

Gaussian Processes (GPs) offer a flexible probabilistic approach to modeling unknown functions. A GP, denoted by $\boldsymbol{\xi} = \left(\xi(\boldsymbol{x})\right)_{\boldsymbol{x} \in \mathcal{X}}$, is a collection of real-valued random variables defined over the same probability space. The defining property of a GP is that for any finite set of inputs $\{\boldsymbol{x}_1, \dots, \boldsymbol{x}_n\} \subset \mathcal{X}$ with $n \geq 1$, the outputs $\xi(\boldsymbol{x}_1), \dots, \xi(\boldsymbol{x}_n)$ form a multivariate Gaussian vector. The GP is fully specified by a mean function $m : \mathcal{X} \to \mathbb{R}, m(\boldsymbol{x}) := \mathbb{E}[\xi(\boldsymbol{x})]$, and a covariance function (kernel) $k : \mathcal{X} \times \mathcal{X} \to \mathbb{R}, k(\boldsymbol{x}, \boldsymbol{x}') := \text{Cov}(\xi(\boldsymbol{x}), \xi(\boldsymbol{x}'))$. \\

In the considered GP modeling context, the target function $f: \mathcal{X} \rightarrow \mathbb{R}$ is treated as a sample from a GP $\boldsymbol{\xi}$, which, in a Bayesian context, can be seen as a prior model. We assume that we observe noisy measurements of $f$ at $n$ input points ${\boldsymbol{x}_1, \dots, \boldsymbol{x}_n}$, resulting in observations of the form $Z_i = f(\boldsymbol{x}_i) + \varepsilon_i$, where the noise terms are i.i.d. with $\varepsilon_i \sim \mathcal{N}(0, \tau^2)$. By conditioning the GP on the observed data $\mathcal{A}_{n} = \{ \boldsymbol{Z}_n = \boldsymbol{z}_n \}$, we obtain a posterior GP characterized by a predictive mean function $m_n(\boldsymbol{x})$ and a posterior covariance function $k_n(\boldsymbol{x}, \boldsymbol{x}')$. We consider the case of ordinary kriging (assuming a constant but unknown mean $m$), for which the formulas of the posterior mean and covariance can be found in \citet{roustant2012dicekriging}. Selecting an appropriate positive definite kernel $k(\boldsymbol{x},\boldsymbol{x'}), \boldsymbol{x}, \boldsymbol{x'} \in \mathcal{X}$ is far from trivial and requires careful consideration, particularly when dealing with non-Euclidean input spaces. In the following section, we discuss ways to design kernels for the G$\times$E product spaces.

\subsection{Kernel combination approaches on product spaces}
\label{sec:kern_comb}

Let us first assume that valid kernels $k_G$ and $k_E$ are given on $\mathcal{X}_G$ and $\mathcal{X}_E$, respectively. We now discuss ways to combine them into valid kernels on $\mathcal{X}_G \times \mathcal{X}_E$. 
Let us denote here by $\xg$ a generic vector of genetic variables in $\mathcal{X}_G$, by $\xe$ a generic vector of environmental variables in $\mathcal{X}_E$, and by $(\xg,\xe)$ a generic (concatenated) vector in the product space $\mathcal{X}_G \times \mathcal{X}_E$. We are thus interested in combining kernels $k_G$ and $k_E$ respectively defined on $\Xg$ and $\Xe$ (that is, with respective input spaces $\Xg^2=\Xg\times \Xg$ and $\Xe^2=\Xe\times \Xe$) into valid kernels $k_{GE}$ on $\Xg \times \Xe$ (thus taking their inputs in $(\Xg\times \Xe)^2=(\Xg\times \Xe)\times (\Xg\times \Xe)$). \\

A first important remark is that $k_G$ and $k_E$ can directly be used as kernels on the product space, in the sense that kernels (informally) defined by $k_{GE}^G((\xg,\xe),(\xg',\xe')=k_{G}(\xg,\xg')$ and $k_{GE}^E((\xg,\xe),(\xg',\xe')= k_{E}(\xe,\xe')$ are both valid on the product space. Either option would amount to ignore part of the variables, which is expected to be detrimental in prediction but would not alter the non-negative definiteness of resulting kernel matrices, thereby delivering valid kernels on $\Xg\times \Xe$. The combination approaches considered in this work essentially revolve around (blockwise) tensor sums and products, in the following sense. First, the tensor sum, 
\begin{equation*}
    k_{GE}^+((\xg,\xe),(\xg',\xe')= k_{G}(\xg,\xg') + k_{E}(\xe,\xe'),
\end{equation*}
defines a valid kernel $\Xg\times \Xe$, following the same mechanism as for usual additive kernels. This operation is immediately extended to non-negatively weighted sums. Beyond this,  
\begin{equation*}
    k_{GE}^{\times}((\xg,\xe),(\xg',\xe')=k_{G}(\xg,\xg') \times k_{E}(\xe,\xe'),
\end{equation*}
is also known to provide valid kernels on $\Xg\times \Xe$, in virtue of the stability of the cone of symmetric non-negative matrices by matrix product. 
Using again summation and non-negative linear combinations, we arrive at a class of kernels synthesizing and extending the latter approaches, whereby $\alpha, \beta,\gamma \geq 0$:
\begin{equation}
\label{Eq_combk}
k_{GE}^{\sim}((\xg,\xe),(\xg',\xe')=\alpha k_{G}(\xg,\xg') + \beta k_{E}(\xe,\xe') + \gamma k_{G}(\xg,\xg') \times k_{E}(\xe,\xe').
\end{equation}

Let us also remark that taking a different $(k_{G}, k_E)$ pair (e.g., with different hyperparameter values) in the sum and product parts of Equation~(\ref{Eq_combk}) still leads to valid kernels on $\Xg\times \Xe$. 

\subsection{Linear mixed-effect models as Gaussian Processes Regression}
\label{lmm_gp}

Before discussing specific kernels, we want to connect the GP modeling framework to the LMM setting. Using LMM notation accounting for G$\times$E interactions and covariates, we can write the observation vector $\boldsymbol{Z}_n$ as \citep{costa2021nonlinear, liu2025incorporating},
\begin{equation*}
\boldsymbol{Z}_n = \mathbf{1}_n m + \boldsymbol{U}_G + \boldsymbol{U}_E + \boldsymbol{U}_{GE} + \boldsymbol{\varepsilon},
\end{equation*}
where $m$ denotes the general mean and $\boldsymbol{\varepsilon} \sim \mathcal{N}(0, \tau^2 \boldsymbol{I}_n )$ a $n$-dimensional noise vector. Furthermore, the random effects of genotype $\boldsymbol{U}_G\sim \mathcal{N}(0, \sigma_G^2 \boldsymbol{Z}_G \boldsymbol{K}_G \boldsymbol{Z}_G^\top )$ and environment $\boldsymbol{U}_E\sim \mathcal{N}(0, \sigma_E^2 \boldsymbol{Z}_E \boldsymbol{K}_E \boldsymbol{Z}_E^\top )$ are considered, where $\sigma_G^2$ and $\sigma_E^2$ denote the variances. The incidence matrices $\boldsymbol{Z}_G$ and $\boldsymbol{Z}_E$ are based on a one-hot encoding of the inputs and $\boldsymbol{K}_{G}$ and $\boldsymbol{K}_{E}$ denote the relationship matrices between the genotypes and the environments, respectively. If the relationship matrices are based on the covariance kernels $k_G$ and $k_E$ applied to the covariates, we can write $\boldsymbol{Z}_E \boldsymbol{K}_E \boldsymbol{Z}_E^\top = [k_E((\boldsymbol{x}_E)_i, (\boldsymbol{x}_E)_j)]_{1 \leq i,j \leq n}$ and $ \boldsymbol{Z}_G \boldsymbol{K}_G \boldsymbol{Z}_G^\top = [k_G((\boldsymbol{x}_G)_i, (\boldsymbol{x}_G)_j)]_{1 \leq i,j \leq n}$. Finally, the interaction term $\boldsymbol{U}_{GE} \sim \mathcal{N}(0, \sigma_{GE}^2 \boldsymbol{Z}_G \boldsymbol{K}_G \boldsymbol{Z}_G^\top \circ \boldsymbol{Z}_E \boldsymbol{K}_E \boldsymbol{Z}_E^\top )$ is modeled using the variance~$\sigma_{GE}^2$, where $\circ$ denotes the Hadamard product of two matrices. By Gaussianity of sums of independent Gaussian-distributed random variables, 
\begin{equation*}
\boldsymbol{Z}_n \sim \mathcal{N} \left(m, \sigma_G^2 \boldsymbol{Z}_G \boldsymbol{K}_G \boldsymbol{Z}_G^\top + \sigma_E^2 \boldsymbol{Z}_E \boldsymbol{K}_E \boldsymbol{Z}_E^\top + \sigma_{GE}^2 \boldsymbol{Z}_G \boldsymbol{K}_G \boldsymbol{Z}_G^\top \circ \boldsymbol{Z}_E \boldsymbol{K}_E \boldsymbol{Z}_E^\top + \tau^2 \boldsymbol{I_n} \right),
\end{equation*}
what coincides with the GP assumption under the combination approach $k_{GE}^{\sim}$ (Eq. \ref{Eq_combk}). \\

To predict at a new input $\boldsymbol{x}$, it is assumed that the response at $\boldsymbol{x}$ follows a joint Gaussian distribution with $\boldsymbol{Z}_n$. 
If $\boldsymbol{x}$ is then conditioned on $\boldsymbol{Z}_n$, we obtain the same posterior Gaussian as with GP modeling \citep[up to some variations due to trend parameters, see][]{roustant2012dicekriging}. In the same way, a LMM approach without G$\times$E interaction could be linked to GP modeling by using the additive kernel combination $k_{GE}^+$ and a LMM with fixed effects could be modeled for instance by using a Universal Kriging approach \citep{matheron1969krigeage, handcock1993bayesian}. However, while LMMs typically rely on frequentist or full Bayesian approaches, GP models employ empirical Bayes and analytical formulas for prediction.

\subsection{Kernels}

Commonly used families of covariance kernels for Euclidean input spaces include the isotropic Matérn kernels, which also cover the isotropic exponential and Gaussian (or square-exponential) kernels as special cases \citep{stein2012interpolation}. In this context, isotropy implies that the kernel values for pairs of locations depend solely on the Euclidean distance between them. For multidimensional inputs, anisotropic versions of these kernels are frequently employed. Here, we consider the isotropic exponential and Gaussian kernel for $\boldsymbol{x},\boldsymbol{x'} \in \mathbb{R}^{\nu}$,
\begin{align*}
    k_{\mathrm{EXP}}(\boldsymbol{x}, \boldsymbol{x'}) = \exp \left( - \frac{||\boldsymbol{x}-\boldsymbol{x'}||}{\theta} \right), \quad
    k_{\mathrm{GAU}}(\boldsymbol{x}, \boldsymbol{x'}) = \exp \left( - \frac{||\boldsymbol{x}-\boldsymbol{x'}||^2}{\theta^2} \right), 
\end{align*}
with $||\boldsymbol{x}-\boldsymbol{x'}|| = \sqrt{(\boldsymbol{x}-\boldsymbol{x'})^\top(\boldsymbol{x}-\boldsymbol{x'})}$ and the correlation length $\theta > 0$. Additionally, we are considering kernels on non-Euclidean spaces, which we discuss in the next sections. 

\subsubsection{Considered kernels on G}

The genetic information $\xg \in \mathcal{X}_G$ considered is given by sequences of strings (SNPs).

\paragraph{Gaussian-GBLUP Kernel ($k_{\mathrm{G: GAU-GBLUP}}$)}
In this approach, the SNP data are numerically encoded based on the major allele frequency. A Gaussian kernel is then applied to the resulting Euclidean distance matrix, following the methodology described by \citet{cuevas2016genomic} and \citet{costa2021nonlinear}. 

\paragraph{Exponential-Hamming Kernel ($k_{\mathrm{G: EXP-HAM}}$)}
The Hamming distance for string sequences $\boldsymbol{x_G},\boldsymbol{x_G'} \in \mathcal{X}_G$ is given by 
$d_{\mathrm{H}}(\boldsymbol{x_G}, \boldsymbol{x_G'}) = \frac{1}{\nu_G}\sum_{i=1}^{\nu_G} \mathbbm{1}(\boldsymbol{x_G}[i]\neq \boldsymbol{x_G'}[i])$ and can be used in the exponential kernel $k_{\mathrm{EXP}}$ instead of the Euclidean distance. The Hamming distance, which counts the number of differing entries between two vectors, can be viewed as the $L_{\infty}$ distance. \citet{hutter2014algorithm} introduced a generalized (weighted) version of this kernel and demonstrated its positive definiteness. While this result applies to the exponential-Hamming kernel, it does not hold for the Gaussian-Hamming kernel  \citep[following a foundational theorem by Schoenberg, see][]{berg1984harmonic}.

\paragraph{Spectrum Kernel ($k_{\mathrm{G: SPE}}$)}
The $k$-spectrum kernel, introduced by \citet{leslie2001spectrum}, captures similarity by comparing substrings of length $k$. The spectrum kernel relies on Mercer's theorem \citep{mercer1909}, defining the kernel as the inner product in a feature space. Therefore, it considers all possible subsequences \(a\) of length \(k\) from the alphabet \(\mathcal{A}\) and defines a feature map from \(\mathcal{X}_G\) to \(\mathbb{R}^{|\mathcal{A}|^k}\) as $\Psi_k(\boldsymbol{x}) = \left(\psi_a(\boldsymbol{x}) \right)_{a \in \mathcal{A}^k}$, where \(\psi_a(\boldsymbol{x})\) represents the frequency of subsequence \(a\) occurring in \(\boldsymbol{x}\). The \(k\)-spectrum kernel is then given by $k_{\mathrm{G: SPE}}(\boldsymbol{x}, \boldsymbol{x'})  = \langle \Psi_k(\boldsymbol{x}), \Psi_k(\boldsymbol{x}') \rangle$. 

\subsubsection{Considered kernels on $E$}

In the considered wheat test case, $\xe \in \mathcal{X}_E$ contains meteorological variables that describe temperature, precipitation, and solar radiation over six time periods. 

\paragraph{Exponential-Euclidean ($k_{\mathrm{E:EXP-EUCL}}$) and Gaussian-Euclidean Kernel ($k_{E:\mathrm{GAU-EUCL}}$)} For these two kernels we apply the exponential and Gaussian kernel to the Euclidean distances of the environmental variables.

\paragraph{Global Alignment Kernel ($k_{\mathrm{E:GAK}}$) }

Global Alignment Kernels (GAK; \citealt{cuturi2007kernel}) compute similarities between time series by building on Dynamic Time Warping (DTW; \citealt{sakoe1970similarity}). Since DTW does not satisfy the triangle inequality, it is strictly not a metric and cannot be used directly as a kernel. As a way around this, \citet{cuturi2007kernel} propose to rely on a soft-minimum, and get the following positive definite kernel,
\begin{align*}
k_{\mathrm{E:GA}}(\boldsymbol{x}, \boldsymbol{x'})  = \sum_{\ell \in \mathcal{L}(\boldsymbol{x}, \boldsymbol{x'})} e^{- \, d_{\mathbf{x}, \mathbf{x'}}(\ell)}, \text{ where } d_{\mathbf{x}, \mathbf{x'}}(\ell) = \sum_{i=1}^{|\ell|} \varphi \left( x_{\ell_1(i)}, x_{\ell_2(i)}' \right),
\end{align*}
using the set of all alignments \(\mathcal{L}(\boldsymbol{x}, \boldsymbol{x'})\) and with the function \(\varphi(\cdot)\) typically chosen as the squared Euclidean distance.

\section{Dataset and Implementation}
\label{sec_dataimple}

\subsection{Wheat dataset}
\label{sec_data}

The wheat data \citep{wheatdata} were collected in multi-environment trials across the Swiss wheat production zone (from Western to Eastern Switzerland, between 390 and 640 MASL), conducted by Agroscope (the Swiss Federal Center for agricultural research) and partners as part of its official Swiss winter wheat variety trials for cultivation and use \citep{strebel2025getreidesorten}. The data includes the performance traits yield (15\% moisture content, average of three replicates, dt per ha) 
and grain protein content (measured on a mixture of three replicates, \%)  per location and year, for a total of 98 varieties. These were grown at 16 locations over the period 1990 to 2023. The management conditions of the trials were low input, i.e. a standard nitrogen fertilization rate and no chemical inputs in the form of growth regulators, insecticides or fungicides. The term ‘environment’ is referring to a combination of trial location and harvest year; we have 263 environments available. \\

Genetic information consists of 12106 SNPs (Single Nucleotide Polymorphisms) marker data. The data were characterized by a SNP call frequency above 80\%, a frequency of homozygous calls above 90\%, and a sample call rate above 90\%. The SNPs are in IUPAC notation with ‘G’, ‘A’, ‘T’ and ‘C’ representing guanine, adenosine, thymine and cytosine bases, and ‘K’, ‘M’, ‘R’ and ‘Y’ representing combinations of these bases as specified in \citet{bowden1985}. For the Gaussian-GBLUP kernel, bi-allelic marker data is employed with homogeneous ‘G = GG’, ‘A =AA’, ‘T=TT’, ‘C=CC’ and heterogeneous ‘K=GT’, ‘M=AC’, ‘R=AG’, ‘Y=CT’. Meteorological data are averaged over six periods covering the Swiss winter wheat growing season: winter (October-February, sowing and vernalization), March (tillering), April (stem elongation), May (heading-flowering), June (flowering and seed filling) and July (harvest). Daily minimum, maximum, and mean temperatures, total and average daily precipitation, and monthly dry days were extracted for each location from the MeteoSwiss Spatial Climate Analysis products (1 km grid; \citealt{meteoswiss_spatial_climate}). Daily solar radiation was obtained from the nearest automated meteorological station \citep{meteoswiss_automatic_network}.

\subsection{Model evaluation}
\label{sec_evaluation}

For evaluation, data are split into training $\boldsymbol{x}_1, \dots, \boldsymbol{x}_n$ and test set $\boldsymbol{x}_{n+1}, \dots, \boldsymbol{x}_m$ and we perform cross-validation on 80 \% of the data using 30 splits. We evaluate predictive performance in two scenarios: new environment and a new variety. Group-based splitting, by environment or variety, prevents information from leaking into the training set. We also consider controlled leakage, allowing one observation in the training set to mimic minimal prior knowledge. For a more detailed discussion of data leakage and its implications in group-based cross-validation, we refer to \citet{guignard2024some}. In agricultural studies, leave-one-out approaches exclude and predict individual environments or varieties \citep[e.g.,][]{tadese2024accuracy}. Our approach extends this by predicting multiple targets per split while keeping the number of training data roughly consistent. \\

To assess the predictive accuracy of the GP, 
we first use the Mean Squared Error (MSE). This metric assesses the mean $m_n(\boldsymbol{x})$ as a predictor, ignoring full probabilistic performance. To capture the probabilistic aspect, the Gaussian predictive distribution must be compared to a single observed value. This can be achieved through the use of a scoring rule \citep{gneiting2007strictly}. Scoring rules assess performance by assigning lower scores to ``better'' probabilistic predictions, ideally capturing both calibration (agreement with observed outcomes) and sharpness (concentration of the predictive distribution). We consider the Continuous Ranked Probability Score (CRPS; \citeauthor{sanders1963subjective}~\citeyear{sanders1963subjective}, \citeauthor{murphy1973new}~\citeyear{murphy1973new}), defined as $\mathrm{CRPS}(F, y) = \int_{-\infty}^{\infty} [F(u) - \mathbbm{1}\{y \leq u\}]^2 \, du$, where $F(\cdot)$ is the cumulative distribution function of the predictive distribution \citep{gneiting2007strictly}.
For a deterministic point predictor, where the predictive distribution reduces to a Dirac delta, the CRPS simplifies to the absolute difference between the predicted and the true value. Finally, we compute the median CRPS for the test set. As a second scoring rule, we consider the log-score \citep[logS; ][]{good1952rational}, $\text{logS}(f,y) = -\log p(y)$, where $p(\cdot)$ is the probability density function of the predictive distribution. Again, we consider the median score over the test set. 

\vspace{1cm}
\subsection{Implementation}
\label{sec_imple}

A key implementation step is estimating the hyperparameters. The covariance matrix of the training data in the noisy case with noise variance $\tau^2$ is given by $\boldsymbol{K}_{obs} = \sigma_K^2 \boldsymbol{K} + \tau^2\boldsymbol{I}_n$, where $\boldsymbol{K}=[k(\boldsymbol{x_i},\boldsymbol{x_j})]_{1 \leq i,j \leq n}$. 
We summarize $\nu= \sigma_K^2+\tau^2$ and denote the proportion explained by the GP as $\varsigma=\sigma_K^2/\left(\tau^2+\sigma_K^2\right).$ 
This gives $\boldsymbol{K}_{obs}=\nu\boldsymbol{K}_\varsigma$, with $\boldsymbol{K}_\varsigma=\varsigma  \boldsymbol{K}+(1-\varsigma)\boldsymbol{I}_n$. The optimal $\nu$, as a function of $(\theta_E,\theta_G,\varsigma)$, can be estimated by likelihood maximization \citep{roustant2012dicekriging}.
The advantage of this approach lies in including the observation noise without increasing the number of hyperparameters, while restricting the search space of $\varsigma$ to $[0,1]$. Recall the kernel sum and product combinations on the product space (Section \ref{sec:kern_comb}). To avoid over-parameterization, we constrain $(\alpha,\beta,\gamma)$ to sum to one. With one hyperparameter per kernel, denoted as $\theta_E$ for the environmental and $\theta_G$ for the genetic kernel, this gives five parameters for $k_{GE}^\sim$, four for $k_{GE}^+$ ($\gamma=0$), and three for $k_{GE}^G$ ($\beta=\gamma=0$), $k_{GE}^E$ ($\alpha=\gamma=0$), and $k_{GE}^\times$ ($\alpha=\beta=0$). \\

Hyperparameters are tuned via maximum likelihood (empirical Bayes). A coarse grid search provides first estimates to initialize the Adam optimizer \citep{Kingma2014AdamAM}, run with a learning rate of 0.01 decayed by 0.8 every 5 steps, for up to 1000 iterations. In order to significantly decrease computation time, we use batching of size 0.5, computing the mean gradient from two randomly selected, disjoint batches, each the size of half the training dataset. For kernels with discrete parameters, we fix them via grid search and furthermore, we restrict the length scales $\theta_G$ and $\theta_E$ to the training set’s maximum point distance. \\

The main parts of the R code used in this study can be found in a GitHub \citep{wheatadvisor2025}. 
We implemented the GP model manually, for the spectrum kernel, we used the \texttt{kernlab} package \citep{kernlab}, and for the global alignment kernel, we used the code available on Marco Cuturi's personal website. In the Hamming distance, we treat missing letters as matches but heterozygotes as separate instances. 
As the meteorological variables have different orders of magnitude, we normalize them within each time period.
\begin{table}[p]
    \centering
     \caption{Considered prediction methods. }
     \vspace{0.3cm}
     \renewcommand{\arraystretch}{1.5} 
    \begin{tabular}{p{1.6cm} | p{1.2cm} | p{6.5cm} | p{5cm}}
        Method & Fixed & Random & Estimation/prediction \\
        \hline 
        GP & $m$ & 
        GP$^G$: $k_{GE}^G$ \newline
        GP$^E$: $k_{GE}^E$ \newline 
        GP$^+$: $k_{GE}^+$ \newline
        GP$^\times$: $k_{GE}^\times$ \newline GP$^\sim$: $k_{GE}^\sim$ \vspace{0.25cm} \newline 
        GP$_1$: $k_{\mathrm{E:GAU-EUCL}}, k_{\mathrm{G:GAU-GBLUP}}$ \newline
        GP$_2$: $k_{\mathrm{E:GAU-EUCL}}, k_{\mathrm{G:EXP-HAM}}$ \newline
        GP$_3$: $k_{\mathrm{E:GAU-EUCL}}, k_{\mathrm{G:SPE}}$ \vspace{0.25cm} \newline
        GP$_4$: $k_{\mathrm{E:EXP-EUCL}}, k_{\mathrm{G:GAU-GBLUP}}$ \newline
        GP$_5$: $k_{\mathrm{E:EXP-EUCL}}, k_{\mathrm{G:EXP-HAM}}$ \newline
        GP$_6$: $k_{\mathrm{E:EXP-EUCL}}, k_{\mathrm{G:SPE}}$ \vspace{0.25cm} \newline
        GP$_7$: $k_{\mathrm{E:GAK}}, k_{\mathrm{G:GAU-GBLUP}}$ \newline
        GP$_{8}$: $k_{\mathrm{E:GAK}}, k_{\mathrm{G:EXP-HAM}}$ \newline
        GP$_{9}$: $k_{\mathrm{E:GAK}}, k_{\mathrm{G:SPE}}$ \newline & 
        Empirical Bayes: \vspace{0.25cm} \newline 
        ML for hyperparameters, \newline Analytical for prediction \newline
        (Sct. \ref{GPmodeling}) \\
        \hline
        BGLR & $m$ & 
        BGLR$^{\sim}$: $\boldsymbol{g}_G + \boldsymbol{g}_E + \boldsymbol{g}_{GE}$ \vspace{0.25cm} \newline
        $k_{\mathrm{E:GAU-EUCL}}, \quad  k_{\mathrm{G:GAU-GBLUP}}$ \newline
        & Full Bayes with Gibbs \\
        \hline
        LMM & $\xe$ & 
        Independent group effects \newline
        LMM$_1$: variety \newline
        LMM$_2$: variety and environment
        & Frequentist: \vspace{0.25cm} \newline  
        REML for hyperparam., \newline BLUP for prediction \\
        \hline
    \end{tabular}
    \begin{tabular}{p{1.75cm} | p{1.2cm} | p{6.5cm} | p{5cm}}
        Average & GLO$_A$ & All training data & \\
        & VAR$_A$ & Data from same variety & \\
        & ENV$_A$ & Data from same environment & \\
    \end{tabular}
    \vspace{.5cm}
    \label{tab:methods}
\end{table}

\subsection{Comparison approaches}
\label{sec_comparison}

As stressed in Section \ref{lmm_gp}, a Bayesian LMM with a covariance structure on environments and genotypes is closely linked to a GP model. To underline this, we run the GP with the state-of-the-art kernels using the R-package for Bayesian Generalized Linear Regression \citep[BGLR;][]{perez2014genome}, which performs a fully Bayesian analysis and samples the hyperparameters, effects and predictions with Gibbs. For the hyperparameters of the kernels (length scale), we use the ML fit of the GP model. Furthermore, we consider a traditional (frequentist) LMM, with fixed effects on the environmental covariates and (independent) group random effects on the variety (and environment). Finally, we compare our GP models to baseline averages: the global average, the environmental average (over all points of the same environment), and the variety average (over points of the same variety as the prediction target). These methods are summarized together with the GP approaches in Table~\ref{tab:methods}.

\section{Results}
\label{sec_results}

\subsection{New environment}

In this section, we are interested in predicting for a new environment. We begin by evaluating GP$^\sim$ using the full kernel combination $k_{GE}^{\sim}$ (Eq.~\ref{Eq_combk}) in conjunction with the nine different kernel configurations introduced in Table~\ref{tab:methods}. Table \ref{tab:results1} summarizes the results for yield and protein content by depicting the median MSE, CRPS and logS. First we consider the scenario without leakage (left of vertical bars). For both traits and with respect to all three metrics, the GPs using the exponential-Euclidean kernel for the environmental information (GP$^\sim_4$, GP$^\sim_5$, GP$^\sim_6$) outperform the ones employing the Gaussian-Euclidean (GP$^\sim_1$, GP$^\sim_2$, GP$^\sim_3$) and the GAK-based kernel (GP$^\sim_7$, GP$^\sim_8$, GP$^\sim_9$). Among the models GP$^\sim_4$, GP$^\sim_5$ and GP$^\sim_6$, we observe minimal performance differences between the genetic kernels. When allowing one controlled leakage point per environment (right of bars), the scores improve drastically. In this scenario, there is generally less difference between the kernels, especially for yield prediction. However, regarding protein, the GAK-based kernel for the environmental variables (GP$^\sim_7$, GP$^\sim_8$, GP$^\sim_9$) and also the very first kernel (GP$^\sim_1$) perform worse than the others. \\

\begin{table}[p]
    \centering
    \caption{Model evaluation results for predictions in a new environment without $|$ with one controlled leakage point. Y represents yield (dt per ha) and P represents grain protein content. The analysis is based on 30 different train/test splits and we show the median of the obtained metrics. For the details regarding the methods we refer to Table \ref{tab:methods}. }
    \vspace{0.3cm}
    \footnotesize\renewcommand{\arraystretch}{1.5} 
    \begin{tabular}{l|lll|lll}
         Method  & MSE Y & CRPS Y & logS Y&  MSE P & CRPS P & logS P\\
         \hline
         GP$_1^\sim$ & 127.91 $|$ 43.34 &  6.31 $|$ 3.67 & 3.88 $|$ 3.34 & 1.54 $|$ 0.56 &  0.69 $|$ 0.43 &  1.64 $|$ 1.18 \\ 
         
         GP$_2^\sim$ & 128.22 $|$ 41.82 &  6.35 $|$ 3.58 &  3.88 $|$ 3.31 & 1.54 $|$ 0.57 &  0.69 $|$ 0.43 & 1.63 $|$ 1.19 \\   
         
         GP$_3^\sim$ & 130.10 $|$ 41.59 &  6.32 $|$ 3.58 &  3.89 $|$ 3.30 & 1.53 $|$ 0.57 &  0.69 $|$ 0.43 &  1.64 $|$ 1.17 \\  
         
         GP$_4^\sim$ & 119.72 $|$ 41.14 &  6.09 $|$ 3.55 & 3.83 $|$ 3.30 & 1.43 $|$ 0.55 &  0.67 $|$ 0.42 & 1.60 $|$ 1.17 \\  
         
         GP$_5^\sim$ & 119.77 $|$ 41.09  & 6.09 $|$ 3.54 & 3.83 $|$ 3.29 & 1.43 $|$ 0.55 & 0.67 $|$ 0.42 &  1.60 $|$ 1.16 \\  
         
         GP$_6^\sim$ & 119.09 $|$ 41.16 &  6.10 $|$ 3.54 &  3.83 $|$ 3.29 & 1.43 $|$ 0.55 &  0.67 $|$ 0.42 & 1.60 $|$ 1.15 \\  
         
         GP$_7^\sim$ & 151.40 $|$ 47.20 &  6.95 $|$ 3.76  &  3.94 $|$ 3.39 & 1.74 $|$ 0.60 &  0.75 $|$ 0.44 &  1.70 $|$ 1.22 \\      
         
         GP$_8^\sim$ & 151.43 $|$ 47.27 &  6.95 $|$ 3.76 &  3.94 $|$ 3.39 & 1.75 $|$ 0.60 &  0.75 $|$ 0.44 &  1.70 $|$ 1.20  \\    
         
         GP$_9^\sim$ & 151.70 $|$ 47.48 &  6.96 $|$ 3.75 &  3.93 $|$ 3.37  & 1.75 $|$ 0.60 &  0.75 $|$ 0.44 &  1.70 $|$ 1.18 \\    
         
         \hline
         
         GP$_5^+$ & 125.21 $|$ 45.69 & 6.22 $|$ 3.78& 3.85 $|$ 3.50 & 1.44 $|$ 0.59 &  0.68 $|$ 0.44  & 1.62 $|$ 1.35 \\   
         
         GP$_5^\times$ & 121.12 $|$ 54.71 & 6.16 $|$ 4.12 & 3.97 $|$ 3.42 & 1.48 $|$  0.70&  0.69 $|$  0.47&  1.73 $|$ 1.25\\ 
         
         GP$_5^G$ & 147.51 $|$  149.34 &  8.68 $|$ 8.80& 34.04 $|$ 34.54& 1.72  $|$  1.72& 0.96 $|$  0.95 & 29.33 $|$ 29.44 \\                
         
         GP$_5^E$ & 152.61 $|$ 82.53&  6.90 $|$ 5.32&  3.98 $|$ 3.92& 1.95 $|$  1.41&  0.81  $|$ 0.70& 1.87 $|$ 2.01 \\                       
         
         \hline
         
        GLO$_A$ & 177.91 $|$ 177.75 & 10.48 $|$ 10.45 &   - & 2.34 $|$ 2.35 & 1.21 $|$ 1.21 & - \\
         
         VAR$_A$ & 148.19 $|$ 148.28 &  9.62 $|$ 9.63 &  - &  1.70 $|$ 1.71 & 1.06 $|$ 1.06 & - \\

         ENV$_A$ & ------ $|$ 96.59 & ----- $|$ 9.63 & - & ----- $|$ 1.81 & ----- $|$ 1.06 & - \\
         
         LMM$_1$ & 156.51 $|$ 148.56  &  9.66 $|$ 9.40  &   - &  1.70 $|$ 1.63 & 1.03 $|$ 1.01 &   -\\
         
         LMM$_2$ & 142.66 $|$ 46.05 &  9.34 $|$ 5.14 &   - & 1.66 $|$ 0.61 & 1.02 $|$ 0.61 &  - \\

         BGLR$^\sim$ & 139.95 $|$ 46.94 &  6.56 $|$ 3.81 & 3.93 $|$ 3.53 & 1.69 $|$ 0.60 & 0.73 $|$ 0.45 & 1.71 $|$ 1.38 
    \end{tabular}
    \label{tab:results1}
\end{table}

\begin{figure}[h]
    \centering
    \includegraphics[width=.48\linewidth]{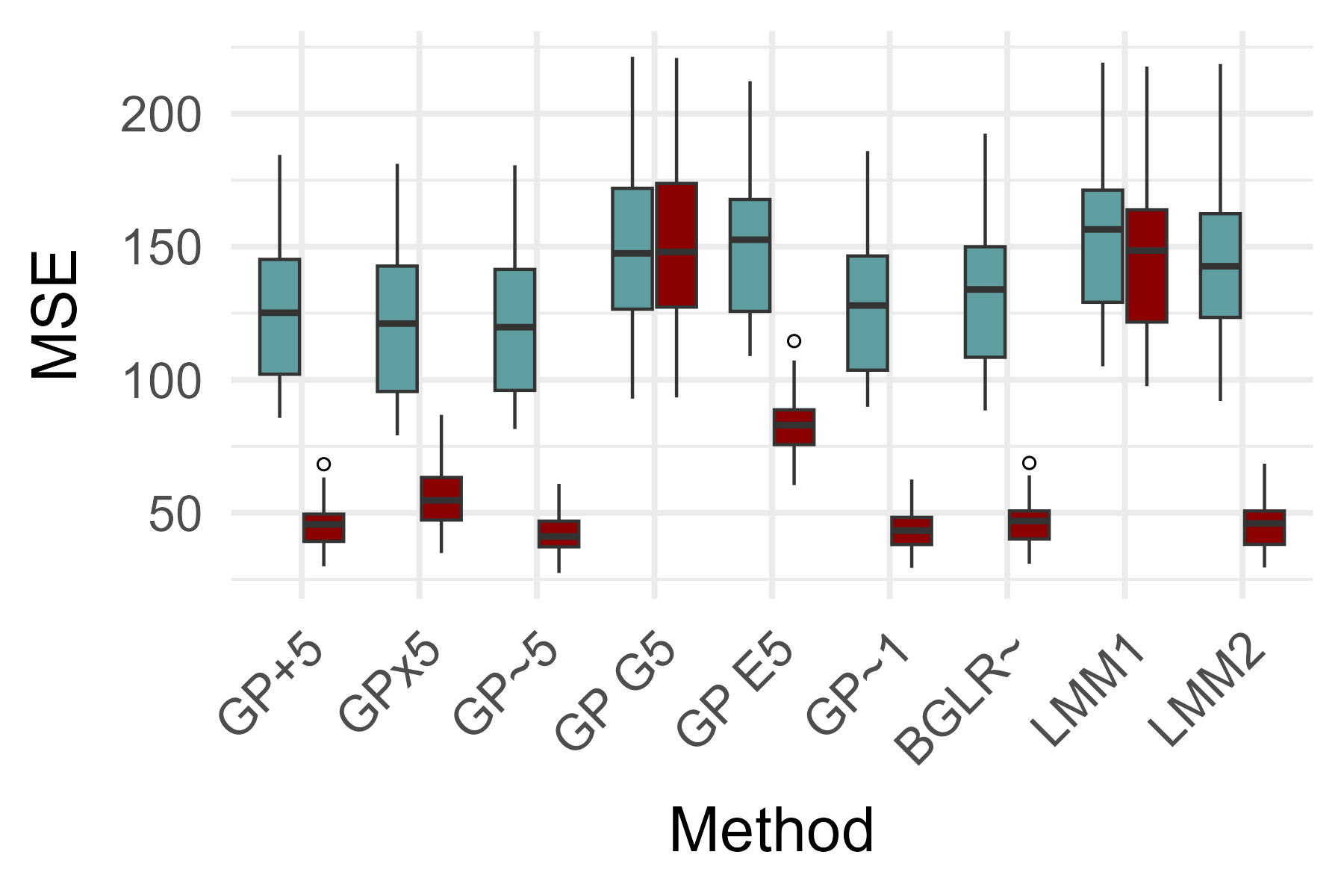}
     \includegraphics[width=.48\linewidth]{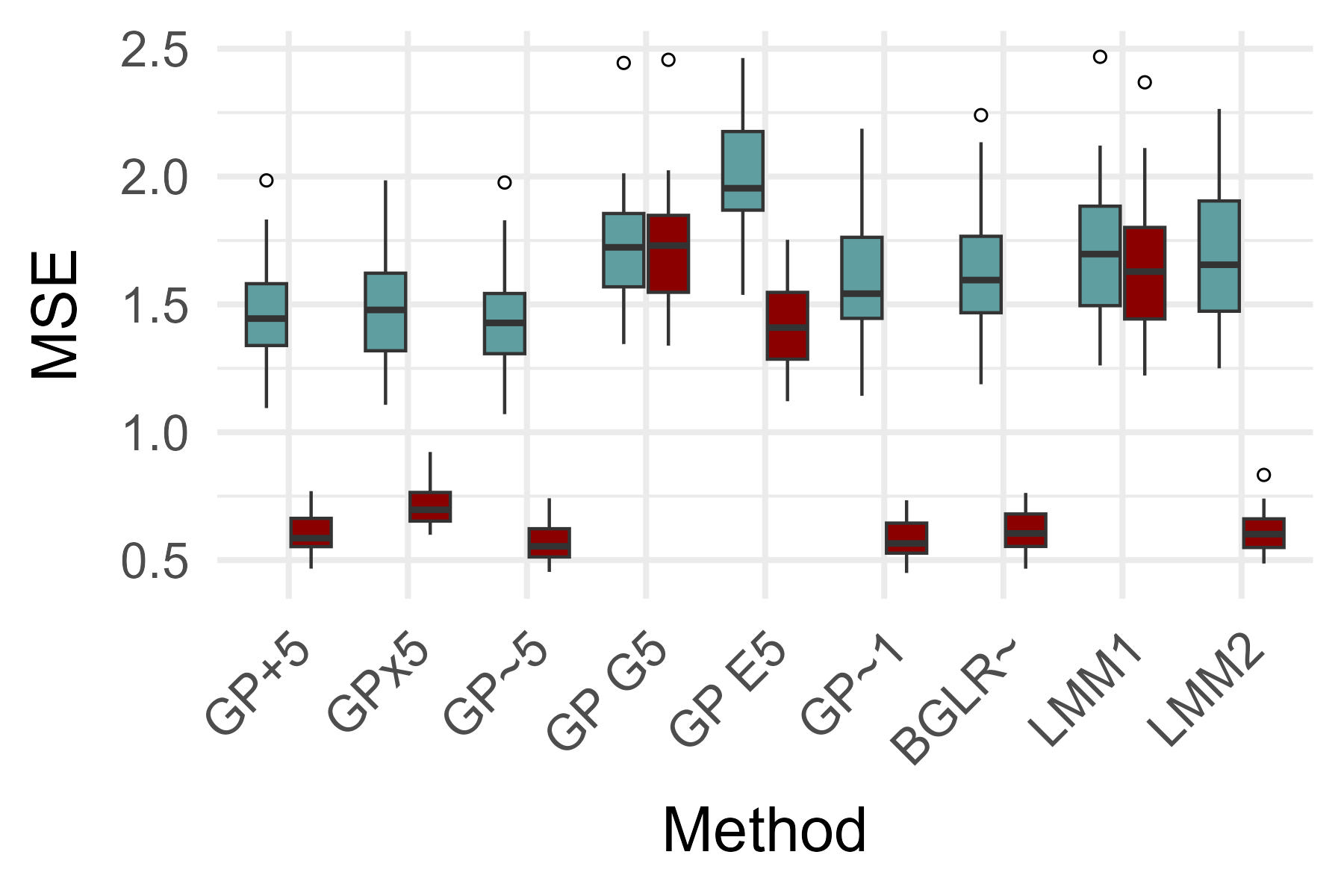} \\
    \hspace{2cm} (a) \hspace{7cm} (b)  
    \includegraphics[width=.48\linewidth]{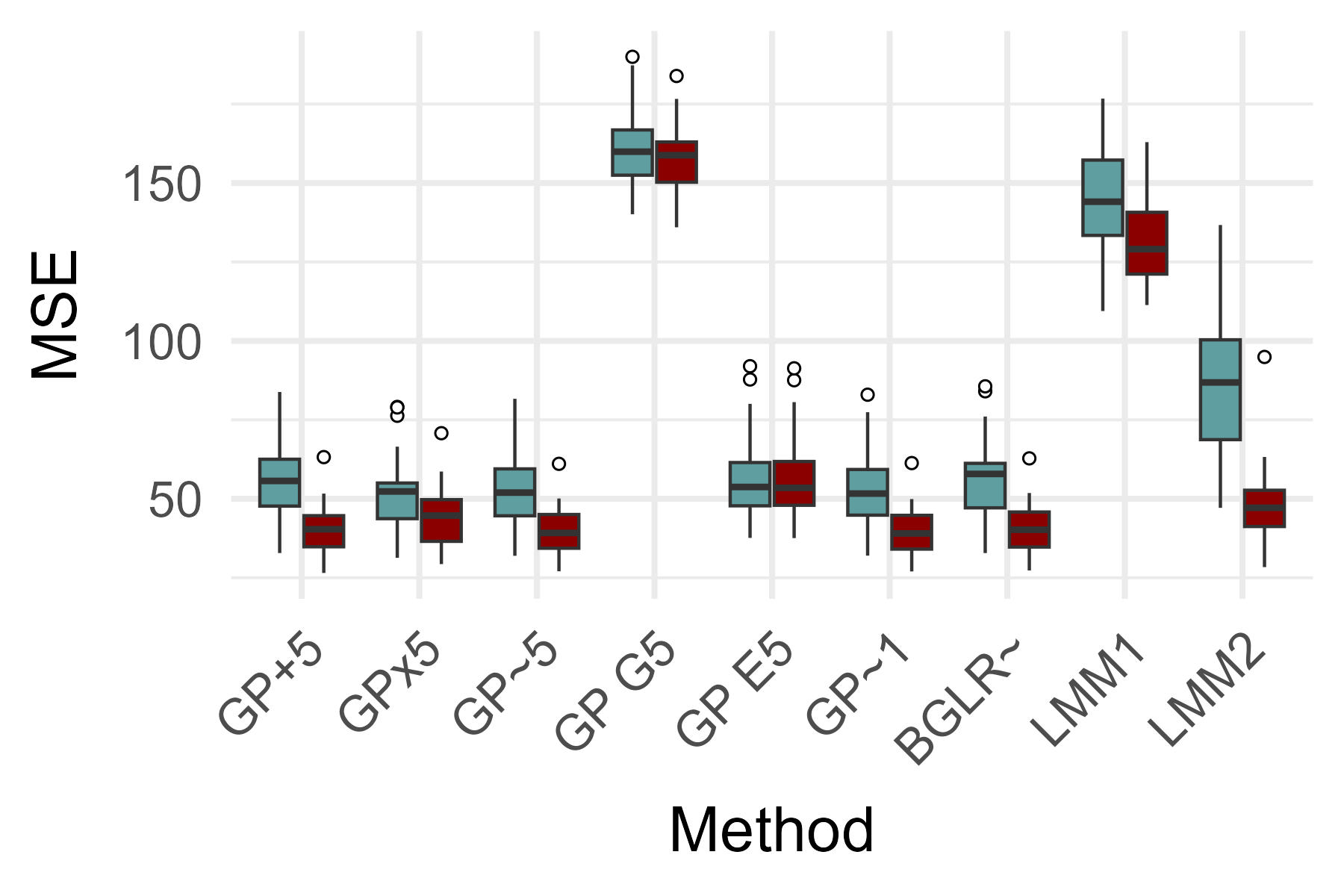} \includegraphics[width=.48\linewidth]{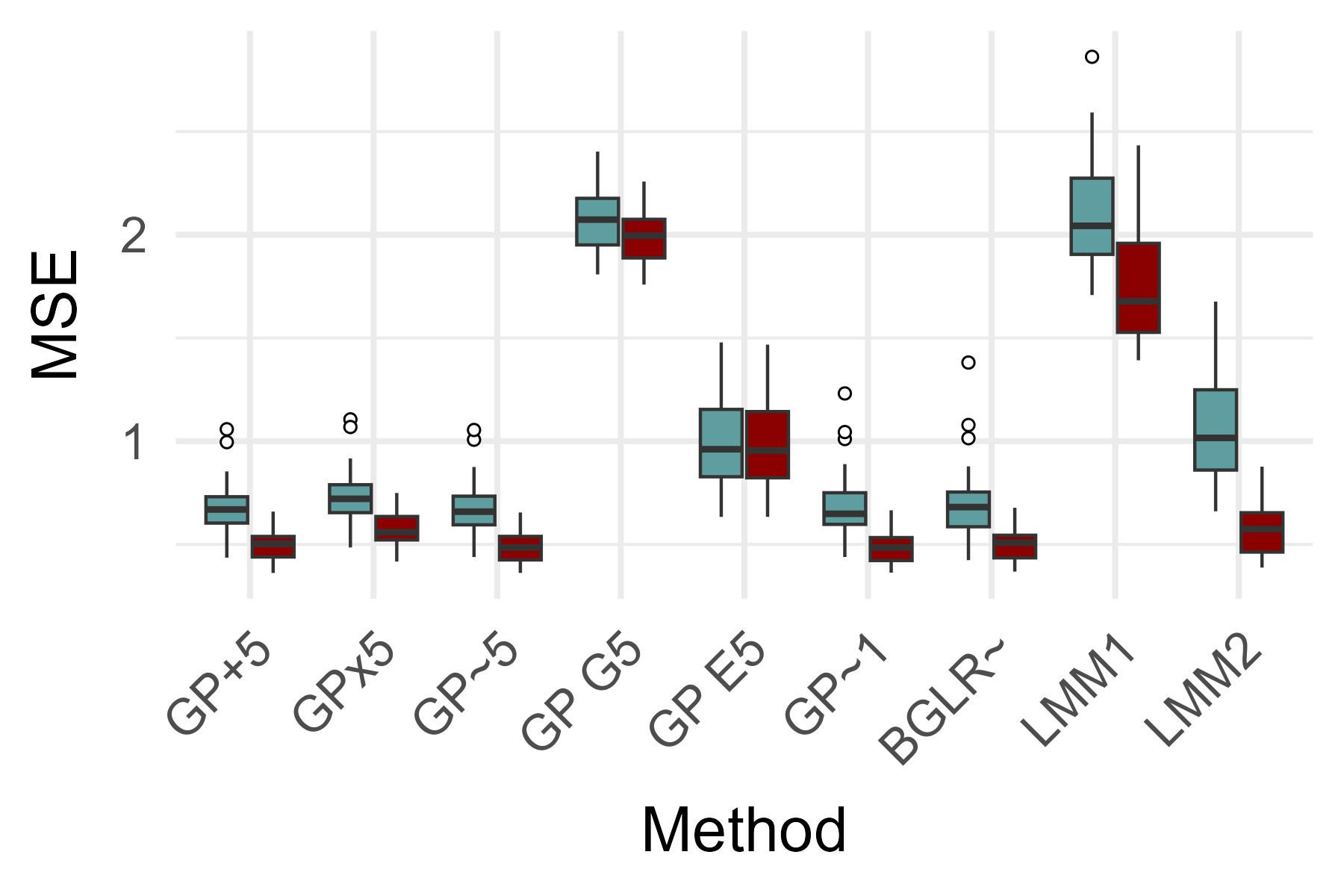}\\
    \hspace{2cm} (c) \hspace{7cm} (d)  
    \caption{Boxplots of the MSE values for prediction of (a,c) yield and (b,d) protein content. The first line (a,b) is for the setting considering a new environment and the second line (c,d) for a new variety. The blue bars depict the values when no leakage is considered, and the red bars the ones where we use one controlled leakage observation per environment or variety. The boxplots show the values across 30 different train/test. }
    \label{fig:results_1}
\end{figure}

Next, we compare the different kernel combination strategies within GP$^G$, GP$^E$, GP$^+$, GP$^\times$, and GP$^\sim$. Thereby, we focus on GP$_5$, which demonstrated convincing performance in the first comparison of the kernels within GP$^\sim$. 
To simplify the interpretation, we complement Table \ref{tab:results1} with boxplots depicting the MSE values of selected methods for protein content and yield prediction (Fig. \ref{fig:results_1}a+b). We observe that GP$^+$, GP$^\times$, and GP$^\sim$ generally exhibit comparable performance, while GP$^G$ and GP$^E$ perform worse than the others. While for protein and without leakage (Fig. \ref{fig:results_1}b, blue bars), the GP using the genetic kernel only (GP$^G_5$) performs clearly better than the one using the environmental kernel only (GP$^E_5$), for yield (Fig. \ref{fig:results_1}a, blue bars) the performances are similar. When considering a single controlled leakage observation of the target environment (red bars), the performance of the environmental kernel only (GP$^E_5$) improves and becomes clearly better than that of the genetic only, even more so for yield than for protein content. Also the medians in CRPS and logS of Table \ref{tab:results1} show very high scores for the GP using the genetic kernel only (GP$^G_5$). The GP using the combinations of kernels still performs better than GP$^E_5$ and for the leakage case, we notice that GP$^+$ and GP$^\sim$ perform better than GP$^\times$. \\ 

\begin{figure}
    \centering
    \includegraphics[width=\linewidth]{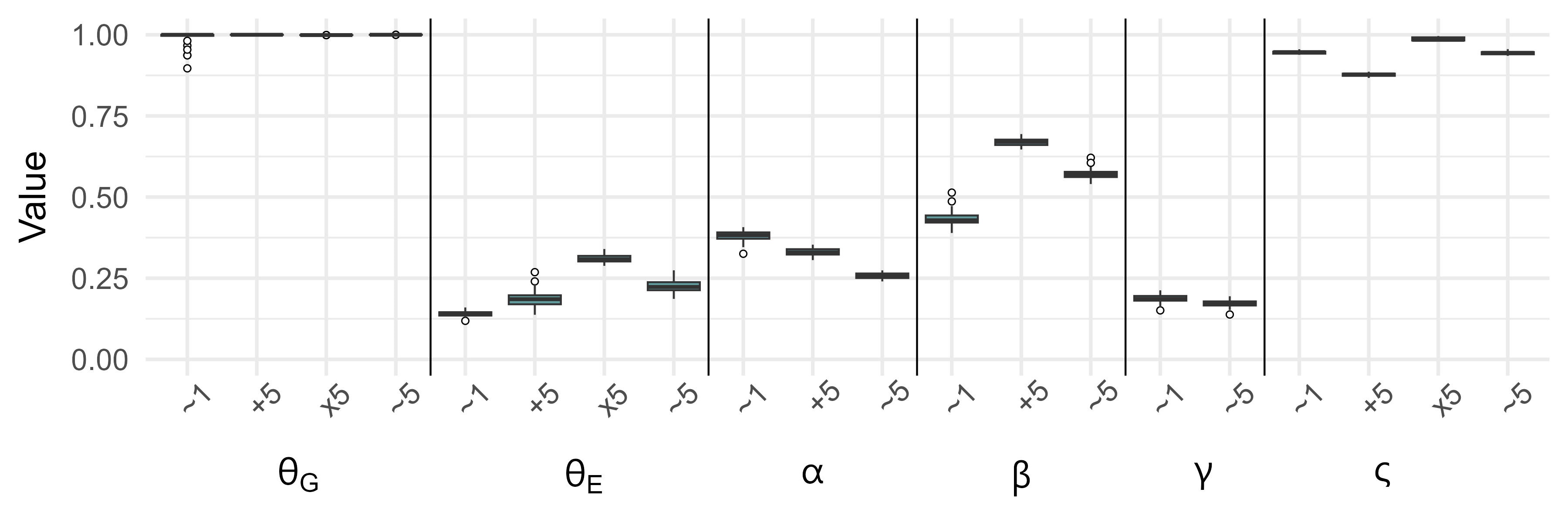}
    \caption{Estimated hyperparameters for the different GP models, fitted using the Adam optimizer. The boxplot shows the values across 30 different train/test splits for a new environment, without leakage, and for the trait yield.}
    \label{fig:hyper_1}
\end{figure}

\pagebreak
To gain a deeper understanding of the inner workings of the GP approaches, Figure~\ref{fig:hyper_1} presents a boxplot of the estimated hyperparameters for the different kernel combinations. We only show the hyperparameters for the scenario targeting yield without leakage, but the results are very similar for protein content and do not drastically change when adding one leakage observation. We immediately observe that $\theta_G$ is always estimated at the boundary value of one, suggesting that the ML approach aims to flatten the effect of the genetic kernels. In contrast, the length scale $\theta_E$ of the environmental kernel is consistently estimated around 0.25 for all GP$^+$, GP$^\times$, and GP$^\sim$ models with kernels 1 and 5. The coefficients for the combinations in GP$^+_5$ assign approximately $\alpha=0.65$ of the weight to the environmental kernel. In the full model, both GP$^\sim_1$ and GP$^\sim_5$ allocate about $\gamma=0.2$ of the weight to the kernel product. Specifically, GP$^\sim_1$ assigns roughly $\alpha=0.35$ to the genetic kernel and $\beta=0.45$ to the environmental kernel, while GP$^\sim_5$ assigns about $\alpha=0.25$ to the genetic kernel and $\alpha=0.55$ to the environmental kernel. Finally, the proportion of variance explained by the GP, denoted $\varsigma$, is estimated similarly across all methods. For GP$_5$, it is lowest for GP$^+$, followed by GP$^\sim$, and highest for GP$^\times$. \\

Figure~\ref{fig:results_1} and Table \ref{tab:results1} also show the scores of the considered state-of-the-art methods: BGLR$^\sim$, \text{LMM}$_1$ and \text{LMM}$_2$. As expected, GP$^\sim_1$ and BGLR$^\sim$ yield similar results for both traits and leakage scenarios. Also, LMM$_1$ (without a random effect for the environment) generally performs worse than most methods, particularly in the leakage scenario. In the controlled leakage situation, the MSE performance of LMM$_2$ (including a random effect for the environment) improves substantially, though it still falls slightly short of the best GP models. Note that the LMMs traditionally do not account for uncertainty, so the logS is missing, and the CRPS values based on the MAE are worse than those of models that explicitly model uncertainty. Finally, we consider the averaging approaches. The global average is clearly outperformed by all methods. The variety and environmental averages perform better, but still clearly worse than the best GP approaches. Interestingly, the variety average performs about as well as the GP$^G_5$ approaches, indicating a lack of predictive power in the genetic kernels. Conversely, the GP$^E$ approach outperforms the environmental average, suggesting more effective kernel choices. 

\subsection{New variety}

In this section, we consider predicting the performance of a new variety. Table~\ref{tab:results2} and Figure~\ref{fig:results_1}c+d report the same assessment metrics as in the previous section. First, we observe that performance in the no-leakage case is substantially better than in the setting targeting a new environment. The difference between leakage and no leakage is less pronounced here, although for the LMM models it still has the largest impact. The scores for GP$^\sim$ using the nine kernels in Table~\ref{tab:results2} show that, in this setting, the spectrum kernel for genotype (GP$^\sim_3$, GP$^\sim_6$, GP$^\sim_9$) performs worse than both the Gaussian-GBLUP (GP$^\sim_1$, GP$^\sim_4$, GP$^\sim_7$) and the exponential-Hamming (GP$^\sim_2$, GP$^\sim_5$, GP$^\sim_8$). Here, the choice of environmental kernel makes little difference. Once again, the differences between the kernels diminish under leakage. \\

We also compare again GP$^G$, GP$^E$, GP$^+$, GP$^\times$, and GP$^\sim$, focusing on GP$_5$. Figure~\ref{fig:results_1}c+d show little difference between the combinations GP$^+$, GP$^\times$, and GP$^\sim$. For yield, GP$^E$ performs remarkably well, and also for protein it still performs reasonably. GP$^G$ performs very poorly in this setting for both traits, likely due to the lack of training data from the same variety. As before and as expected, GP$^\sim_1$ and BGLR$^\sim$ produce very similar results. LMM$_1$ again cannot compete, while LMM$_2$ performs well in the leakage scenario. Finally, considering the averaging approaches in Table \ref{tab:results2}, the global average performs poorly here as well. The environmental average performs similarly to the GP$^E$ approaches when not accounting for uncertainty quantification (MSE). In the leakage case, the variety average relies on one single measurement of the same variety and performs very poorly.

\begin{table}[p]
    \centering
    \caption{Model evaluation results for predictions for a new variety without $|$ with one controlled leakage point. Y represents yield (dt per ha) and P represents grain protein content. The analysis is based on 30 different train/test splits and we show the median of the obtained metrics. The methods are detailed in Table \ref{tab:methods}. }
    
    \vspace{0.3cm}
    \footnotesize\renewcommand{\arraystretch}{1.5} 
    \begin{tabular}{l|lll|lll}
         Method  & MSE Y & CRPS Y & logS Y&  MSE P & CRPS P & logS P\\
         \hline
         GP$_1^\sim$ & 51.66 $|$ 39.04 & 3.99 $|$ 3.46 & 3.42 $|$ 3.30 & 0.65 $|$ 0.48 & 0.45 $|$ 0.39 & 1.23 $|$ 1.10 \\
         
         GP$_2^\sim$ & 52.42 $|$ 39.48 & 4.01 $|$ 3.47 & 3.42 $|$ 3.28 & 0.65 $|$ 0.48 & 0.46 $|$ 0.39 & 1.22 $|$ 1.08 \\
         
         GP$_3^\sim$ & 62.37 $|$ 43.17 & 4.35 $|$ 3.64 & 3.48 $|$ 3.33 & 0.74 $|$ 0.49 & 0.49 $|$ 0.4 & 1.27 $|$ 1.08 \\
         
         GP$_4^\sim$ & 52.53 $|$ 38.99 & 4.02 $|$ 3.45 & 3.42 $|$ 3.3 & 0.66 $|$ 0.49 & 0.45 $|$ 0.39 & 1.22 $|$ 1.10 \\ 
         
         GP$_5^\sim$ & 51.96 $|$ 39.19 & 4.03 $|$ 3.46 & 3.41 $|$ 3.28 & 0.66 $|$ 0.48 & 0.46 $|$ 0.39 & 1.21 $|$ 1.09 \\
         
         GP$_6^\sim$ & 61.48 $|$ 41.24 & 4.39 $|$ 3.59 & 3.49 $|$ 3.33 & 0.73 $|$ 0.5 & 0.48 $|$ 0.4 & 1.26 $|$ 1.09 \\
         
         GP$_7^\sim$ & 51.5 $|$ 38.96 & 3.98 $|$ 3.47 & 3.44 $|$ 3.35 & 0.66 $|$ 0.48 & 0.46 $|$ 0.4 & 1.24 $|$ 1.13 \\   
         
         GP$_8^\sim$ & 53.21 $|$ 39.6 & 4.03 $|$ 3.48 & 3.44 $|$ 3.34 & 0.65 $|$ 0.48 & 0.46 $|$ 0.39 & 1.22 $|$ 1.10 \\
         
         GP$_9^\sim$ & 62.32 $|$ 43.7 & 4.42 $|$ 3.66 & 3.5 $|$ 3.33 & 0.76 $|$ 0.5 & 0.49 $|$ 0.4 & 1.27 $|$ 1.10 \\ 
         
         \hline
         
         GP$_5^+$ & 55.67 $|$ 40.36 & 4.16 $|$ 3.61 & 3.62 $|$ 3.69 & 0.67 $|$ 0.5 & 0.46 $|$ 0.41 & 1.35 $|$ 1.33 \\ 
         
         GP$_5^\times$ & 52.33 $|$ 44.69 & 4.00 $|$ 3.70 & 3.44 $|$ 3.38 & 0.72 $|$ 0.56 & 0.48 $|$ 0.42 & 1.28 $|$ 1.16 \\

         
         GP$_5^G$ & 159.91 $|$ 158.79 & 7.92 $|$ 7.98 & 6.42 $|$ 6.83 & 2.07 $|$ 2 & 0.9 $|$ 0.89 & 3.33 $|$ 3.85 \\

         
         GP$_5^E$ & 53.75 $|$ 53.48 & 4.77 $|$ 4.76 & 7.86 $|$ 7.89 & 0.96 $|$ 0.95 & 0.65 $|$ 0.65 & 6.39 $|$ 6.48 \\                     
         
         \hline
         
        GLO$_A$ & 177.71 $|$ 177.77 & 10.48 $|$ 10.49 & -  & 2.40 $|$ 2.40 & 1.25 $|$ 1.25 & - \\
         
         VAR$_A$ & ------ $|$ 257.13 & ------ $|$ 12.57 & - & ------ $|$ 3.07 & ------ $|$ 1.41 & - \\

         ENV$_A$ & 53.73 $|$ 53.21 & 5.76 $|$ 5.76 & - & 0.94 $|$ 0.94 & 0.77 $|$ 0.77 & - \\
         
         LMM$_1$ & 144.07 $|$ 129.05 & 9.33 $|$ 8.81 & - & 2.04 $|$ 1.68 & 1.14 $|$ 1.02 & - \\
         
         LMM$_2$ & 86.85 $|$ 47.16 & 7.42 $|$ 5.36 & - & 1.02 $|$ 0.57 & 0.83 $|$ 0.60 & - \\
         
         BGLR$^\sim$ & 57.88 $|$ 40.22 & 4.23 $|$ 3.61 & 3.66 $|$ 3.68 & 0.68 $|$ 0.51 & 0.48 $|$ 0.41 & 1.40 $|$ 1.35 \\
    \end{tabular}
    \label{tab:results2}
\end{table}

\section{Discussion and Conclusions}
\label{sec_discu}

Optimizing multi-environment trials and recommending varieties based on the local environment are two very common themes in agricultural research with significant potential for application. This paper introduces a GP modeling approach for G$\times$E prediction, and links it to state-of-the-art Bayesian LMM methods. Thereby, we explore the impact of kernel choice and kernel combination, and confirm that GP models and Bayesian LMMs operate in essentially the same way. In doing so, we not only test the Gaussian kernels already used, but also alternative kernels specifically designed for the structure of the data at hand. \\

There is considerable scope to discuss and improve the choice of kernels within the GP modeling.
We observed that the performance of the state-of-the-art approaches (BGLR$^\sim$ and GP$_1^\sim$) could be improved using alternative kernels, primarily when targeting new environments (Tab.~\ref{tab:results1}). However, improvements did not come from the more complex kernels for time series and strings, but rather from the simpler exponential kernel approaches (GP$_5^\sim$). In particular, the time series kernel based on global alignment performed poorly when predicting for a new environment (Tab.~\ref{tab:results1}). This may be because, for wheat performance, not only the sequence of weather events matters, but also their specific timing within the growing cycle. 
Similar behavior was observed for the spectrum kernel for strings when predicting a new variety (Tab. \ref{tab:results2}), suggesting the use of alternative kernels. The spectrum kernel considered here is also quite simplistic and has been extended, for instance, by \citet{eskin2002mismatch}. \\

We observe that the kernel combination approaches (GP$^+$, GP$^\times$, and GP$^\sim$) overall did not lead to substantial differences. Only in the leakage case did the additive model GP$^+$ perform worse than the others. Across all settings, the full model GP$^\sim$ consistently performed slightly better. In general, the genotype kernels perform poorly when predicting new varieties. As shown in Fig. \ref{fig:results_1}c and d, the environmental-only model GP$^E$ is not substantially improved by adding genotype information in GP$^\sim$, while the genotype-only model GP$^G$ performs almost as poorly as the global average (Tab.~\ref{tab:results2}). Also in the setting concerned with a new environment, the length scale $\theta_G$ is maximized to flatten out the genetic effect as much as possible (Fig.~\ref{fig:hyper_1}), and the variety average performs about as well as GP$^G$. This performance, especially in the scenario predicting for a new variety, indicates that the considered genetic kernels effectively only ‘work’ for the same variety, meaning that their measure of genetic similarity is not very effective in-between varieties. This clearly highlights the potential of exploring alternative kernels beyond the Gaussian-GBLUP and our proposed kernels. \\

The proposed environmental kernel $k_{\mathrm{E:EXP-EUCL}}$ performs better, and in Figure~\ref{fig:results_1}a and~b, which target a new environment, using a kernel combination clearly outperforms using a single kernel. However, including one leakage observation from the same environment substantially improves prediction accuracy, highlighting the importance of site- and year-specific management and soil characteristics for wheat performance. This effect of local information is also evident in the stronger prediction results for new varieties (Table~\ref{tab:results2}), compared to prediction for new environments (Table~\ref{tab:results1}). In this setting, all available local environmental data can be leveraged, providing insights into local management and soil characteristics, which leads to good prediction performance even without variety leakage. To mitigate this effect, incorporating additional environmental covariates describing soil and management practices, as suggested by \citet{buntaran2021projecting}, could be beneficial. Extending the GP with kernels for these inputs enables a comprehensive model integrating genetic, environmental (weather and soil) and management information. \\  

We hope this paper serves as a useful proof of concept demonstrating that the GP framework, with its flexibility for handling diverse input spaces, provides a promising foundation for improved G$\times$E predictions. 
We see various potential applications: 
Local variety recommendations at the farm level with the GP model enabling the inclusion of new varieties from the official Swiss recommended list \citep{strebel2025getreidesorten}, investigation of future weather scenarios during the winter wheat growing season and improvement of multi-environment trials.  
Moreover, GPs enable joint modeling of yield and protein content, allowing for local analysis of their relationship.
Finally, GP modeling opens the door to sequential design strategies \citep{chevalier2014fast}, and, as it inherently provides uncertainty quantification, to risk-averse strategies and decision-aid tools. 

\paragraph{Data and Code} The anonymized wheat dataset used in this work is available on Zenodo \citep{wheatdata}. 
Key code is available on a Github repository (\href{https://anonymous.4open.science/r/Wheat_advisor_anonymous-E380/}{link}).

\paragraph{Acknowledgements} 
Lea Friedli acknowledges support by the Swiss National Science Foundation (grant number: 225353). Tim Steinert and David Ginsbourger acknowledge the support of the Digitization Commission (DigiK) of the University of Bern via the project “Perception in Statistics, Econometrics and Probability”. Part of this research was performed while Tim Steinert was visiting the Institute for Mathematical and Statistical Innovation (IMSI), supported by the National Science Foundation (Grant No. DMS-1929348). David Ginsbourger would like to thank the Isaac Newton Institute for Mathematical Sciences, Cambridge, for support and hospitality during the program Representing, calibrating \& leveraging prediction uncertainty from statistics to machine learning, where work on this paper was undertaken that was partially supported by EPSRC grant EP/Z000580/1 and by a grant from the Simons Foundation. Nathalie Wuyts, Lilia Levy H\"aner, Didier Pellet and Juan M. Herrera acknowledge support by Agroscope, swiss granum, the Swiss Federal Office for Agriculture [project 'Wheat Advisor', grant no. 19.03], the Schweizerischer Getreideproduzentenverband, Prometerre, Fresh Food \& Beverage Group and Timac Agro Swiss. Computations were performed on UBELIX (https://www.id.unibe.ch/hpc), the HPC cluster at the University of Bern

\bibliography{bib}

\end{document}